\documentclass[12pt,a4paper]{article}

\title{Macroscopic analysis and modelling of\newline multi-class, flexible-lane traffic}

\usepackage{graphicx,fancyhdr}  
\usepackage{amsmath,times}
\usepackage{subfigure}
\makeatletter

\renewcommand{\@thesubfigure}{(\alph{subfigure})\hskip\subfiglabelskip}
\renewcommand{\@@thesubfigure}{\alph{subfigure}}
\makeatother
\usepackage{latexsym}
\usepackage{color}

\newcounter{remark}

\providecommand{\dfrac}[2]{{\displaystyle\frac{#1}{#2}}\:}

\usepackage[sort,numbers]{natbib}
\newcommand{\citeh}{\citep}
\newcommand{\citev}{\cite}
\newcommand{\trbcite}[1]{\citev}

\usepackage[pagewise]{lineno}
\setcitestyle{authoryear}

\usepackage[blocks]{authblk}

\newcommand{\insti}{Delft University of Technology,  Transport \& Planning\authorcr  Stevinweg 1,  Delft, The Netherlands}

\newcommand{\fnamei}{Knoop}
\newcommand{\conti}{+31 15 278 8413 \authorcr v.l.knoop@tudelft.nl}
\author{V. L. {\fnamei}  PhD}
\affil{\insti \authorcr \conti}

\author{W.J. Schakel PhD}
\affil{\insti \authorcr w.j.schakel@tudelft.nl}

\author{T.P. van Oijen MSc}
\affil{\insti \authorcr t.p.vanoijen@tudelft.nl}

\author{prof. L. Leclercq PhD}
\affil{Univ. Gustave Eiffel -- Lyon\authorcr ludovic.leclercq@entpe.fr}

\newcommand{\authorfamilinames}{{Knoop, Schakel, Van Oijen, and Leclercq}\\} 

\usepackage{graphicx,fullpage,amsmath,subfigure,natbib,times}

\newcommand{\halvepagina}{0.5\textwidth}
\newcommand{\derdepagina}{0.32\textwidth}

\newcommand{\fig}{figure~}
\newcommand{\Fig}{Figure~}
\newcommand{\eq}{equation~}

\newcommand{\dens}{k}
\newcommand{\denstot}{\dens_\textrm{tot}}
\newcommand{\denseff}{\dens_\textrm{eff}}

\begin{document}
\maketitle
\maketitle
\newpage
\section*{abstract}

An excessive demand of vehicles to a motorway bottleneck leads to traffic jams. Motorbikes are narrow and can drive next to each other in a lane, or in-between lanes in low speeds. This paper analyses the resulting traffic characteristics and presents numerical scheme for a macroscopic traffic flow model for these two classes. The behavior included is as follows. If there are two motorbikes behind each other, they can travel next to each other in one lane, occupying the space of one car. Also, at low speeds of car traffic, they can go in between the main lanes, creating a so-called filtering lane. 
The paper numerically derives functions of class-specific speeds as function of the density of both classes, incorporating flexible lane usage dependent on the speed. The roadway capacity as function of the motorbike fraction is derived, which interesting can be in different types of phases (with motorbikes at higher speeds or not). We also present a numerical scheme to analyse the dynamics of this multi-class system. We apply the model to an example case, revealing the properties of the traffic stream , queue dynamics and class specific travel times. The model can help in showing the relative advantage in travel time of switching to a motorbike.

\vspace{2cm}\noindent 
\textbf{Key words:} Traffic flow theory, multi lane traffic, multi class traffic, macroscopic traffic modelling, cell transmission model
\newpage

\section{Introduction}
In freeway traffic, there are various types of vehicles. Most freeway traffic is lane bound: passenger cars and trucks mostly stay behind each other within their lane (or change lanes to another lane. On a microscopic scale (i.e., at the level of the vehicle), this behaviour is hence modelled often as car-following (see e.g. \citeh{Tre:2000,New:2002}), combined with a lane change model (e.g., MOBIL \citeh{Kes:2007a} or LMRS \citeh{Sch:2012a}). 
Motorbikes have a different type of behavior. They can follow the traffic stream in case of high speeds. However, when the speeds are low, they can go in-between two lanes, a phenomenon called filtering, studied by real-world trajectories for instance by \citev{Lee:2012}. Moreover, if there are two motorbikes, they can drive next to each other in one lane. On motorways, the nature of traffic is hence in-between lane-bound and two dimensional flows. The microscopic processes can hence qualitatively be understood. In this paper, we address how motorbikes with that dual nature character can influence travel times. To do so, we aim to reconstruct the aggregate dynamics of a combined traffic stream with traditional approaches like a fundamental diagram and density dependent evolution of the traffic stream.

The generic idea of motorbikes is that they can reduce travel times. This is because indeed they can drive next to each other, hence effectively doubling the roadway capacity if only motorbikes are present. Moreover, they might in low speeds create an additional lane in between the lanes. So even if the (assumed) headway-speed relationship might be the same for motorbikes as for cars, the effective road capacity can be higher for motorbikes. It is unknown, however, which traffic states will emerge at capacity and which capacities and travel times will result from a mix of cars and motorbikes. 

Traditionally, traffic engineers have solved multi-class traffic so with a passenger car equivalent. This tells how for how many passenger cars a particular vehicle type counts in terms of capacity. For instance, the Dutch Highway Capacity Manual does provide a number for trucks, stating that a truck ``consumes'' the same amount of capacity as 2.0 passenger cars \citeh{RWS:2015}. Several studies have considered the PCE factor for motorbikes or motorized two-wheelers in urban environments (e.g., ; this is for instance mainly in India with urban traffic with many motorized two wheelers a major step. A more advanced model tries to capture the dynamical nature of the space a motorbike takes. For instance, \cite{Nair:2011} models the flow as porous flow, where motorbikes can go through the gaps at lower speeds. This is validated by \citev{Amb:2014}. This has been further formalized and mathematically analyzed by \citev{Gas:2018}. Note they explicitly take into account speed functions for both classes which depend on the densities of both classes. 

For two-wheelers and motorcars have also been analysed to have independent densities. This continuous approach does not take into account the dual nature of the motorbike behaviour at motorways: being bound to main lanes sometimes, and sometimes being able to use an addtional filtering lane. 

Other approaches merely exploit the passenger car equivalents. The idea is that traffic can be split into classes, each with their own density. The concept started by \citev{Hoo:1999} is that these densities then can be combined into one aggregated density. \citev{Won:2002a} show that this can be made into a model with different vehicle classes, which operate at different speeds for the same aggregated density. Based on this, for instance, \citeh{Ngo:2007} presents a model where speeds in congestion are the same and speeds in free flow are different. \citev{Cha:2003} present a case where they implement variations in vehicle lengths. This is also the basis for the model presented by \cite{Lint:2008}, further developed by \citeh{Wag:2014}, which come up with speed-dependent passenger car equivalent factors. A clear overview of the various preceding works is given in \citev{Log:2008}, which themselves present a combination where vehicle classes have different speeds, and some can be in congestion and some not. The above works all have in common that the main idea is that there is one main aggregated density which determines the traffic states for both classes. 

\citev{Wie:2019} presents a model for bicyclists and cars. Their model is also based on independent densities for cars and bicycles. Depending on the densities, cars can pass cyclists (low densities) or cyclists can pass cars (when cars are stopped, for instance). Whereas not as rigorous with lane bound traffic, their main line of thought is can be applied to the problem at hand here. The model they develop, however, is formulated in Lagrangian coordinates \citeh{Lec:2007-1, Lav:2003}. This means that the speed of travellers is calculated and their positions are updated. It is harder to analyse typical bottlenecks in road networks, which are usually positioned in particular locations. For this, a discretisation scheme fixed in space would be preferred.

There are various ways of discretizing traffic streams, of which the Cell Transmission Model \citeh{Dag:1994} is probably the most well known. In this case, the road is split into cells of fixed length, and the flow between the cells is calculated, which yields new densities. The densities then in turn determine the new flow. This have lead to multi-lane extensions, of which \citev{Nag:2019} present a recent example. \citev{Lav:2006} introduced a reduction of flow as function of lane changes, caused by voids due to bounded acceleration. Several models include these concepts into macroscopic models, e.g. \citeh{Lec:2016, Jin:2018}. These models are limited to applying to a case with a fixed number of lanes. 

For multi-class modelling, we should also look into the work of \citev{Dag:2002}. He proposed that there can be two phases of traffic. One in which all the lanes travel at the same speed, a so-called one-pipe regime, and one in which some lanes travel at higher speeds than other lanes, a two-pipe regime. In his paper, he points at speed and lane choices for drivers. In the current paper, these assumptions will not be used. However, we will refer to ``one pipe regime'' for a situation where all traffic moves at the same speed and ``two pipe regime'' for a situation where one traffic stream (the motorbikes in a filter lane) travel at a different speed than another traffic stream (the cars in the main lanes).

All in all, the gap we work on in this paper is (i) qualitative equilibrium description of traffic with switching lane discipline (sometimes a filtering lane, sometimes not), and (ii) a macroscopic traffic operations in Eulerian coordinates which is able to describe traffic dynamics for a traffic with switching lane discipline. 

The approach we have in the remainder of the paper is as follows. We first discuss the traffic behaviour and resulting traffic states in a static way in section \ref{sec_FDandcapacity}. That section also analyses the capacity as function of the fraction of motorbikes. Then, this is incorporated in a dynamic model. Section \ref{sec_dynamics} presents this dynamic model, and presents the application thereof to an example case. Also some properties of dynamic states are being shown there. The last section presents the discussion and conclusions.

\section{Fundamental diagrams and capacities}\label{sec_FDandcapacity}
The study does the following assumptions:
\begin{itemize}
\item Traffic operations of cars and motorbikes both follow the same triangular fundamental diagram if driving in line behind each other.
\item If the speed of the main lane traffic drops to under 50 km/h, motorbikes can move in between. The speed difference is at maximum 20 km/h (safety constraint, in line with Dutch traffic guidelines). 
\item Motorbikes will take the fastest lane (i.e., they will not remain in the filtering lane when the speed of the filtering lane is lower than in the main lanes).
\item If two motorbikes travel one behind another in one of the main lanes, they can drive next to each other, hence occupying only one ``spot'' and needing only one headway to the predecessor. Note this is a binary event: the motorbikes either are behind each other in the same line or not. 
\end{itemize}
The goal of this section is to show fundamental diagrams, in this case speed for a class of vehicles as function of density of both classes. Graphically they are represented as surfaces (one surface for each class). As it turns out, there are various distinct traffic phases or regimes. For reasons of simplicity, the remainder of this section will first derive the phase diagrams and fundamental diagrams for the case that motorbikes cannot drive next to each other. Then, the problem is further expanded by including the fact that motorbikes can drive next to each other. Finally, we will derive the capacity of the combined section. To make the work more tangible, we will explain and compute the situations based on a situation of a 3 lane road, where a 4th lane of motorbikes can go in-between. The principles are equally applicable to roads with another number of lanes. 

\subsection{Phase diagrams and fundamental diagrams}
\begin{figure}
\subfigure[Phase 1: low speeds, high motorbike fractions, so motorbike traffic in an additional filtering lane and in between the main travel lanes as well; filtering lane at the same speed as main travel lane]{\includegraphics[width=\textwidth]{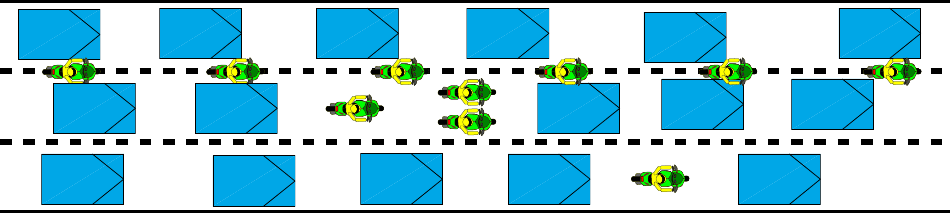}}\\
\subfigure[Phase 2: high speeds, so motorbike traffic in the main travel lanes and all travelling at the same speed]{\includegraphics[width=\textwidth]{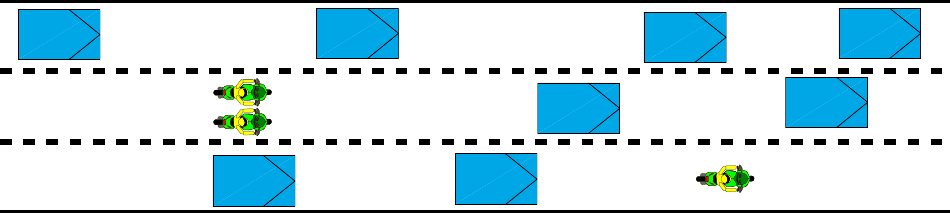}}\\
\subfigure[Phase 3: low speeds, low motorbike fractions, so motorbike traffic in an additional filtering lane travelling at a higher speed than the car traffic in the main travel lanes]{\includegraphics[width=\textwidth]{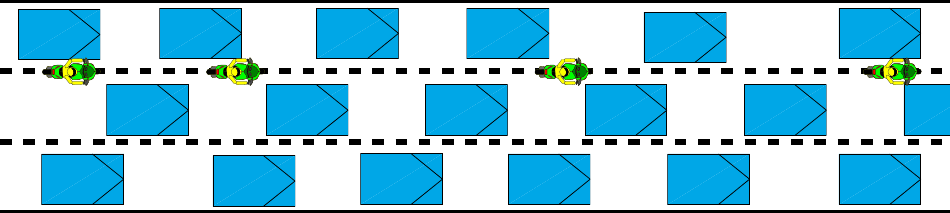}}
\caption{The 3 phases}\label{fig_phases_topview}
\end{figure}

\begin{figure}
\includegraphics[width=\halvepagina]{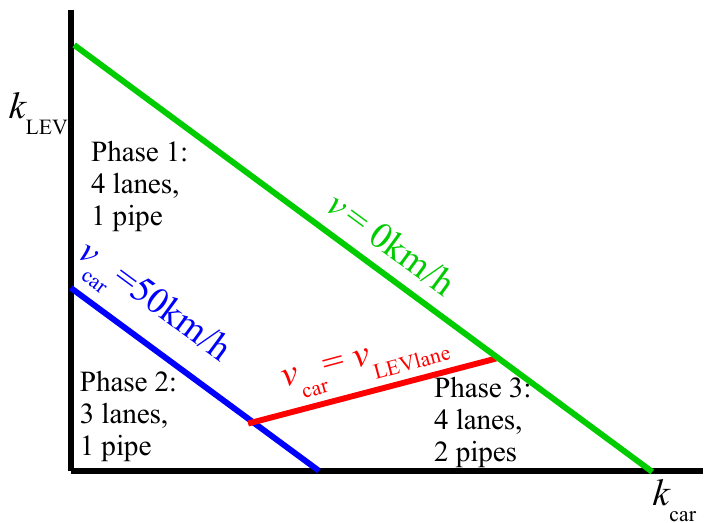}
\caption{The various regions in the density-density plane}\label{fig_phases}
\end{figure}

In this section we will first indicate the various phases that can exist in equilibrium. Figure \ref{fig_phases} shows the various phases possible, and \fig \ref{fig_phases} shows them in a phase diagram, as function of the independent variables, being the density of cars and the density of motorbikes. After we have discussed the phases, we derive the speeds for the respective classes in each of the phases. 

Recall that in this section we assume that motorbikes cannot drive next to each other in one lane, only at speeds lower than 50 km/h, they might go in between lanes. Hence, for combined densities which cause the speed in the main lanes to be exceeding 50 km/h, cars and motorbikes are in the main lanes. This is called Phase 2 in the phase diagram in \fig \ref{fig_phases}. The combined density at which this phase ends is from the regular fundamental diagram where the speed is equal to 50 km/h. For higher densities, the road changes from a 3 lane road into a 4 lane road, with (some) motorbikes in the additional lane. 

The remaining traffic conditions can be separated in two phases. Phase 3 is the phase where motorbikes in the additional lane actually go quicker than cars (and perhaps some motorbikes) in the main lane. Phase 1 is the phase where all speeds are equal. Intuitively, the difference is understood. If the amount of motorbikes is low, they can go in between. For higher fraction of motorbikes, they are too numerous to all fit in the filtering lane and will also contribute to lower speeds in the main lanes. The speed at which they do not any more have benefit from the fitering stops if the speed in the filtering lane is equal to the speed in the main lanes. At the red line indicating the difference between the phases, the speeds are hence equal.

We can explicitly find this line by the following procedure. For all combinations of densities in phase 1 and 3, we can determine the speed of the filtering lane if all motorbikes would drive there. We can also find the speed for the remaining lanes if there are only cars. If the speed in the filtering lane is lower, traffic will be mixed (phase 1) and otherwise, it will not (phase 3). 

We now have the phases in the phase diagram. To find the class-specific speeds, we do the following, per phase:
\begin{enumerate}
\item In phase 1, we have all 4 lanes operating at the same speed. The combined density of cars and motorbikes can hence be distributed over 4 lanes, and the speed can be read from the fundamental diagram. The speed is the same for cars and motorbikes. 
\item In phase 2, all traffic is mixed in a 1 pipe regime, over 3 lanes. We determine the combined density which is distributed over 3 lanes, and with the regular fundamental diagram, we find the speed, being the speed for both cars and motorbikes.
\item In phase 3, we find the speed of the 3 main lanes by taking the car density and divide that over 3 lanes. This yields the lane-specific density for which a car speed can be found in the fundamental diagram. The motorbikes are in the filter lane, so their speed is based on the minimum of (1) the speed matching the density of the motorbikes in one lane according to the fundamental diagram and (2) the speed of the cars plus 20 km/h. In this phase, the speed of motorbikes is to a large extent independent on the car density, and vice versa due to the fact that the classes are separated.
\end{enumerate}

\subsection{Adding that motorbikes can drive next to each other in one lane}\label{sec_nexttoeachother}

In previous section, we derived a speed function as function of density. We did not take into account that the motorbikes can drive next to each other. This is what we will introduce in this section.

\subsubsection{Correcting for motorbikes next to each other} 

The previous speed functions were based on how many ``spots'' were occupied on the road. As it turns out now, sometimes there can be two vehicles (i.e., motorbikes) in one spot -- but only if two motorbikes follow each other directly. We will now analytically derive which density of cars and motorbikes (and hence which fraction of motorbikes) will lead to how many occupied spots. 

We represent the demand as an infinite random sequence, sampled from the set $\{ m, c \}$, where $m$ represents a motorbike and $c$ represents a car. The sampling probability of $m$ and $c$ are $P(m)$ and $P(c) = 1-P(m)$ respectively. Vehicles cannot overtake, but two consecutive motorbikes in the sequence are assumed to occupy one single spot if possible. The resulting number of occupied spots per length unit is defined as the density in the remainder of this analysis.

In order to calculate this density, we notice that each infinite random sequence of vehicles can be uniquely written as an infinite sequence of sub-chains, drawn from the set $\{ c, m \rightarrow c, m \rightarrow m \}$, where $m \rightarrow c$ represents a motorbike followed by a car and $m \rightarrow m$ represents a motorbike followed by a motorbike. The special property of this representation of the demand is that the full width of the lane is guaranteed to be occupied after each element in the sequence, which makes that the number of occupied spots of a sub-chain does not depend on previous sub-chains. The sampling probability of sub-chains $c$, $m \rightarrow c$ and $m \rightarrow m$ are respectively $P(c) = 1-P(m)$, $P(m \rightarrow c) = P(m) \cdot (1-P(m))$ and $P(m \rightarrow m) = P(m)^2$. The number of spots that sub-chains $c$, $m \rightarrow c$ and $m \rightarrow m$ occupy, are 1, 2 and 1 respectively.

Now, we select the first $i$ elements from the transformed sequence. The expected number of vehicles in this sequence equals 
\begin{align}
E(N) &= i \cdot (1 \cdot P(c) + 2 \cdot P(m \rightarrow c) + 2 \cdot P(m \rightarrow m)) \\
     &= i \cdot ( (1-P(m)) + 2 \cdot (P(m) \cdot (1-P(m))) + 2 \cdot P(m)^2) \\
     &= i \cdot (1 + P(m))
\end{align}
The expected number of occupied spots of these vehicles equals
\begin{align}
E(L) &= i \cdot (1 \cdot P(c) + 2 \cdot P(m \rightarrow c) + 1 \cdot P(m \rightarrow m)) \\
     &= i \cdot ( (1-P(m)) + 2 \cdot (P(m) \cdot (1-P(m))) + 1 \cdot P(m)^2) \\
     &= i \cdot (1 + P(m) - P(m)^2)
\end{align}

Now, we have the expected number of vehicles and the expected number of slots. Relating them to each other, we get the number of expected slots from the number of expected vehicles.
\begin{equation}
E(L)=E(N)\frac{1+P(m) - P(m)^2 }{1+P(m)} =E(N) \left(1-\frac{P(m)^2}{1+P(m)}\right)
\label{eq_fractionchange}
\end{equation}

In traffic engineering terms, $E(N)$, the expected number of vehicles, is the density, whereas $E(L)$ is the density of occupied slots. The main takeaway from this analysis is hence we can obtain a correction factor to change the (actual) density into a new (effective) density if motorbikes are allowed to drive next to each other. This is found from the ratio $\frac{E(L)}{E(N)} = \left(1-\frac{P(m)^2}{1+P(m)}\right)$. 

\subsubsection{Resulting combined traffic operations}
This means we can redo the steps mentioned in the previous section, but taking into account the changes in density. These correction needs to be taken into account for finding the boundaries and for the speeds in the zones. Moreover, the fraction of traffic which is motorbike is not constant if a part of the motorbikes can use the filtering lane. Therefore, we re-iterate here the steps but now with this distinction.

\begin{figure}
\includegraphics[width=\halvepagina]{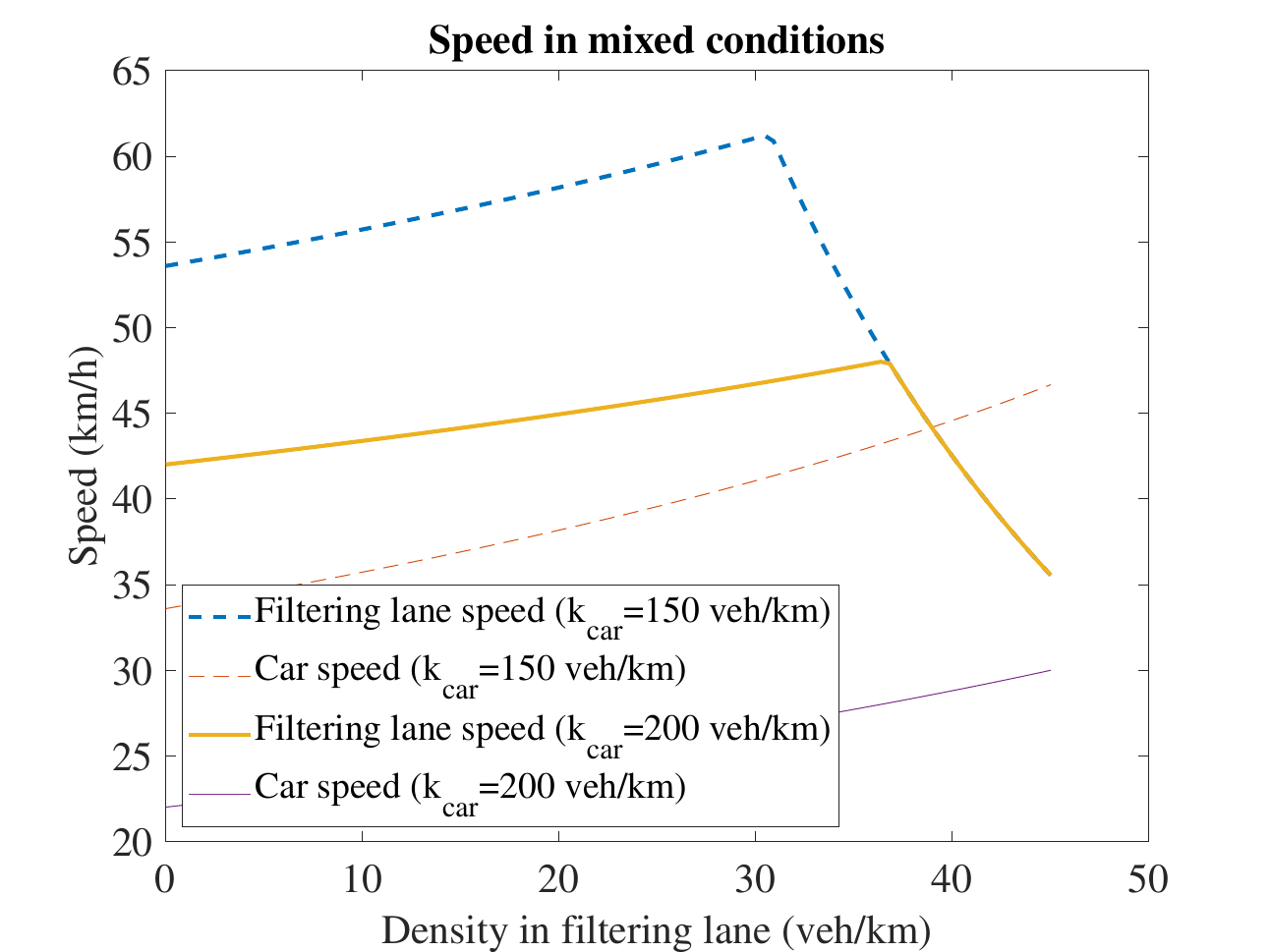}
\caption{The speed in the various lanes as function of the density in the filtering lane.  The case considers a motorbike density of 45 veh/km and a car density of 150 or 200 veh/km. A part of that density is moved to the filtering lane.}\label{fig_speedsfiltering}
\end{figure}

Let's consider again the 3 phases of \fig \ref{fig_phases}. We take the following steps:\begin{enumerate}
\item Consider all combinations of density of cars and motorbikes. Determine a joint density is called $\denstot$. 
\item This can be transformed into an \emph{effictive} density $\denseff$, indicating what the speed would be if there all vehicles would drive in the main driving lanes. This transformation can be done with \eq \ref{eq_fractionchange}. 
\item From the effective density, using the one-class fundamental diagram a speed can be derived.
\item Determine the boundaries between the zone 1 and 2, i.e., where does the traffic drive a lower speed than 50 km/h, based on the effective density. For these areas, the speed is determined based on the single-class FD
\item For the remaining combinations of density, determine for each possible density of motorbikes in the filtering lane:\footnote{this step is illustrated with an example in \fig \ref{fig_speedsfiltering}.}
\begin{enumerate}
\item The speed of the motorbikes in the motorbike lane, based on the single-class single-lane fundamental diagram. This will be decreasing with the number of motorbikes in the motorbike lane for higher densities. For lower densities in the filtering lane and congested conditions in the main lines, the speed is determined by the restriction of 20 km/h higher than the car speed.
\item The density of motorbike traffic in the remaining lanes, and the fraction of motorbike traffic in these remaining lanes
\item The speed of the traffic in the remaining lanes, which will be increasing with the number of motorbikes in the filtering lane, accounting for the correction based on the motorbike fraction, equation \ref{eq_fractionchange}.
\item The traffic state in the filtering lane. This is found by either (a) the main lanes cannot handle the remaining traffic even if the filtering lane is filled till jam density. In this case, the state is not possible (inaccessible region, right top of \fig \ref{fig_phases}. (b) a density for the filtering lane exist which gives the filtering lane the same speed as the remaining lanes. In that case, we are in the 1 pipe regime. (c) a state exist which even if all motorbikes are in the filtering lane, they have a higher speed than if they were joining the main line. In that case, we are in a 2 pipe regime where motorbikes use the filtering lane. The traffic state in the main lines are determined by the car; the traffic density in the filtering lane is determined by all the motorbikes, and the speed in the filtering lane is the maximum of the speed according to the 1 lane fundamental diagram with motorbike density and the car speed plus 20 km/h. Note that this procedure implicitly separates the 1 and 2 pipe regimes. 
\end{enumerate}
\end{enumerate}

\begin{figure}
\subfigure{\includegraphics[width=\halvepagina]{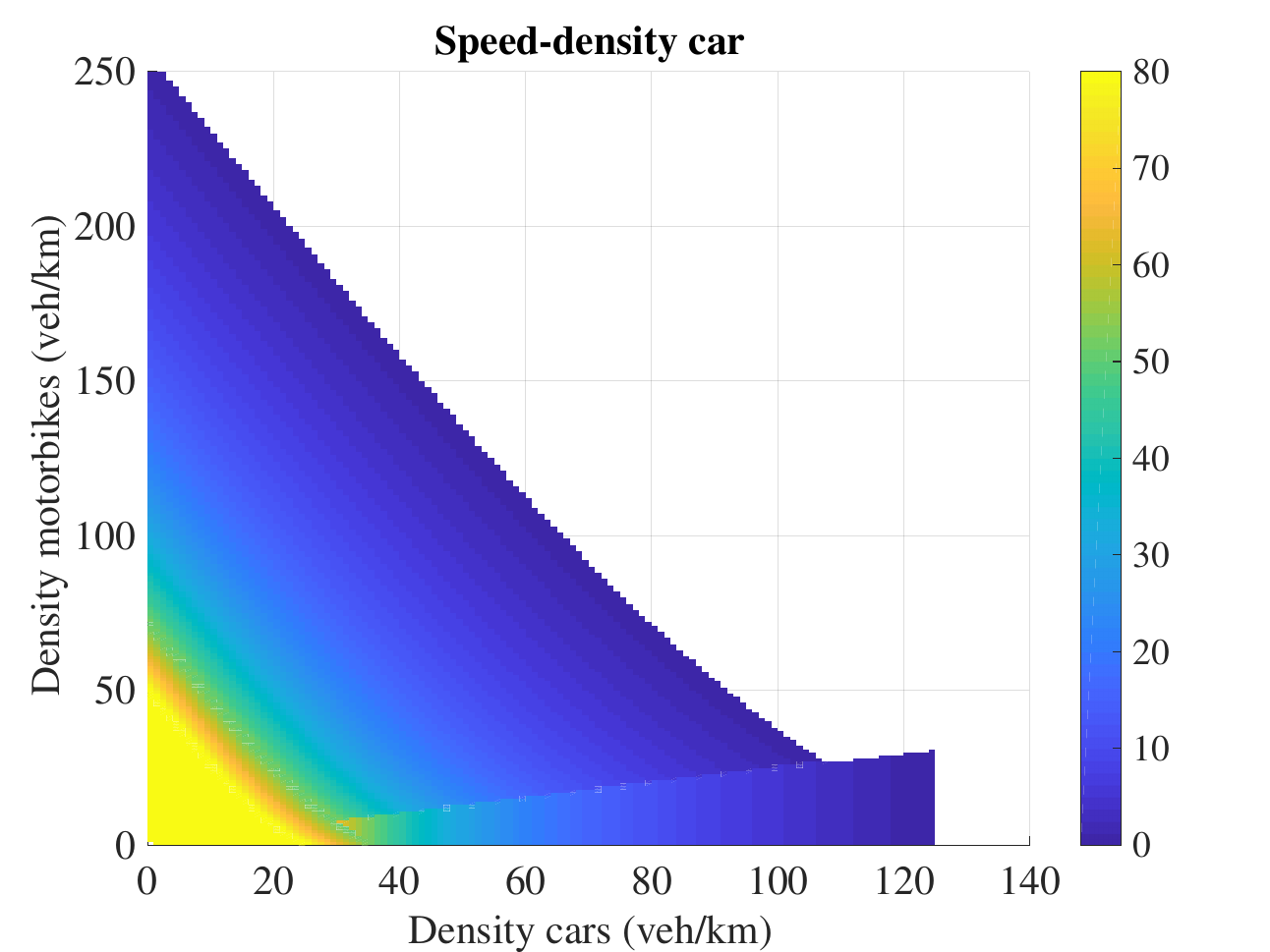}}
\subfigure{\includegraphics[width=\halvepagina]{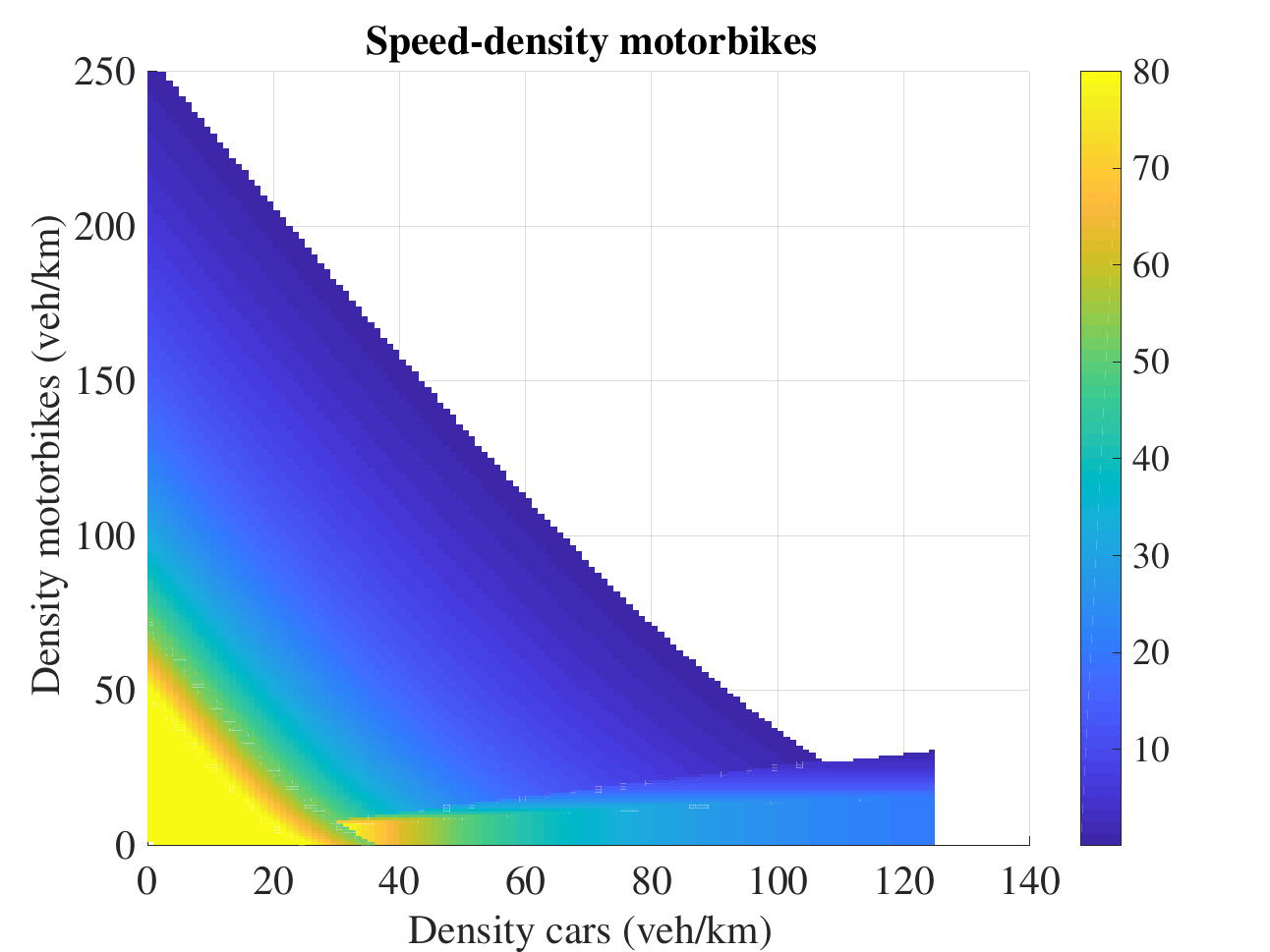}}\\
\subfigure{\includegraphics[width=\halvepagina]{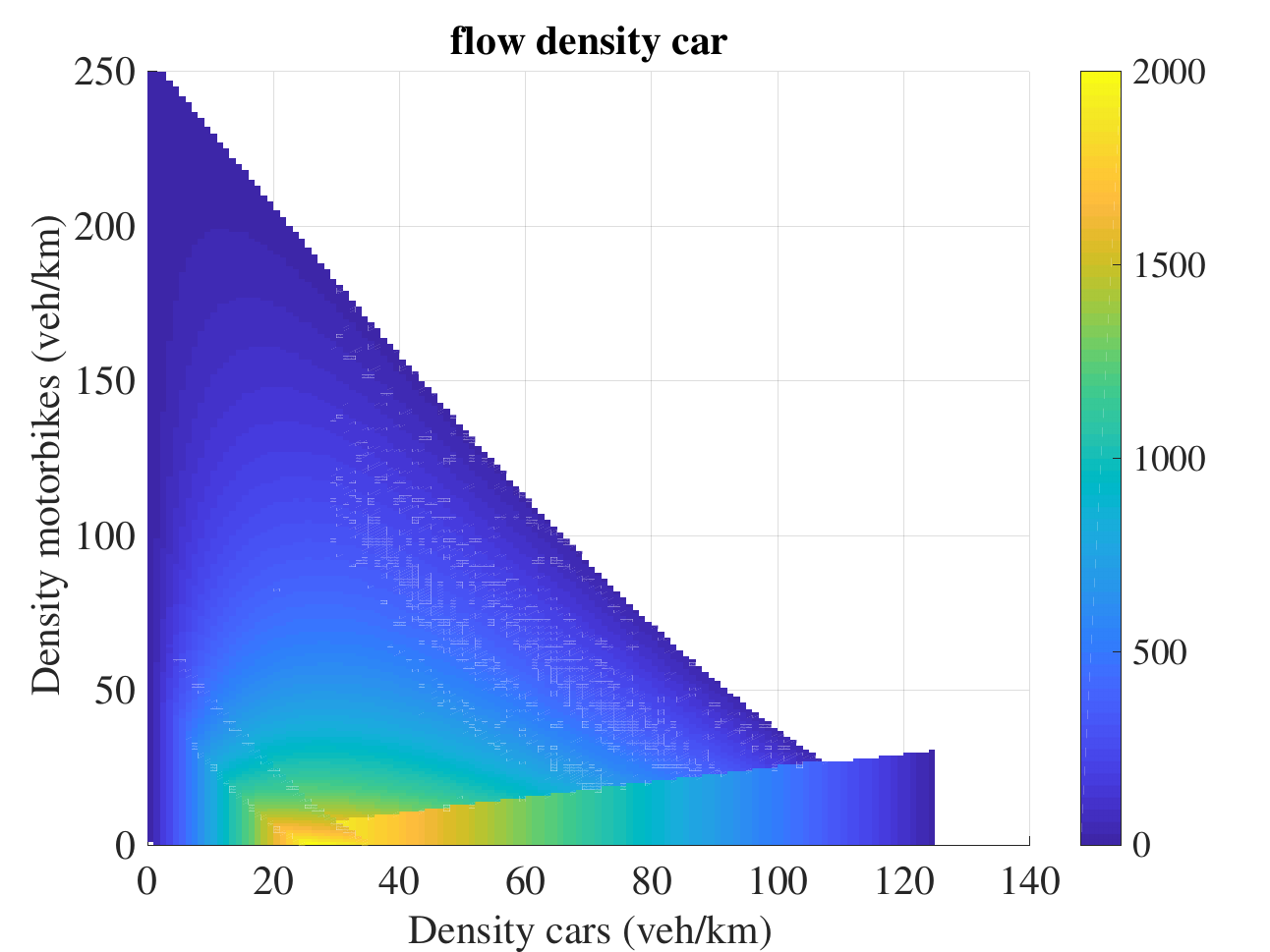}}
\subfigure{\includegraphics[width=\halvepagina]{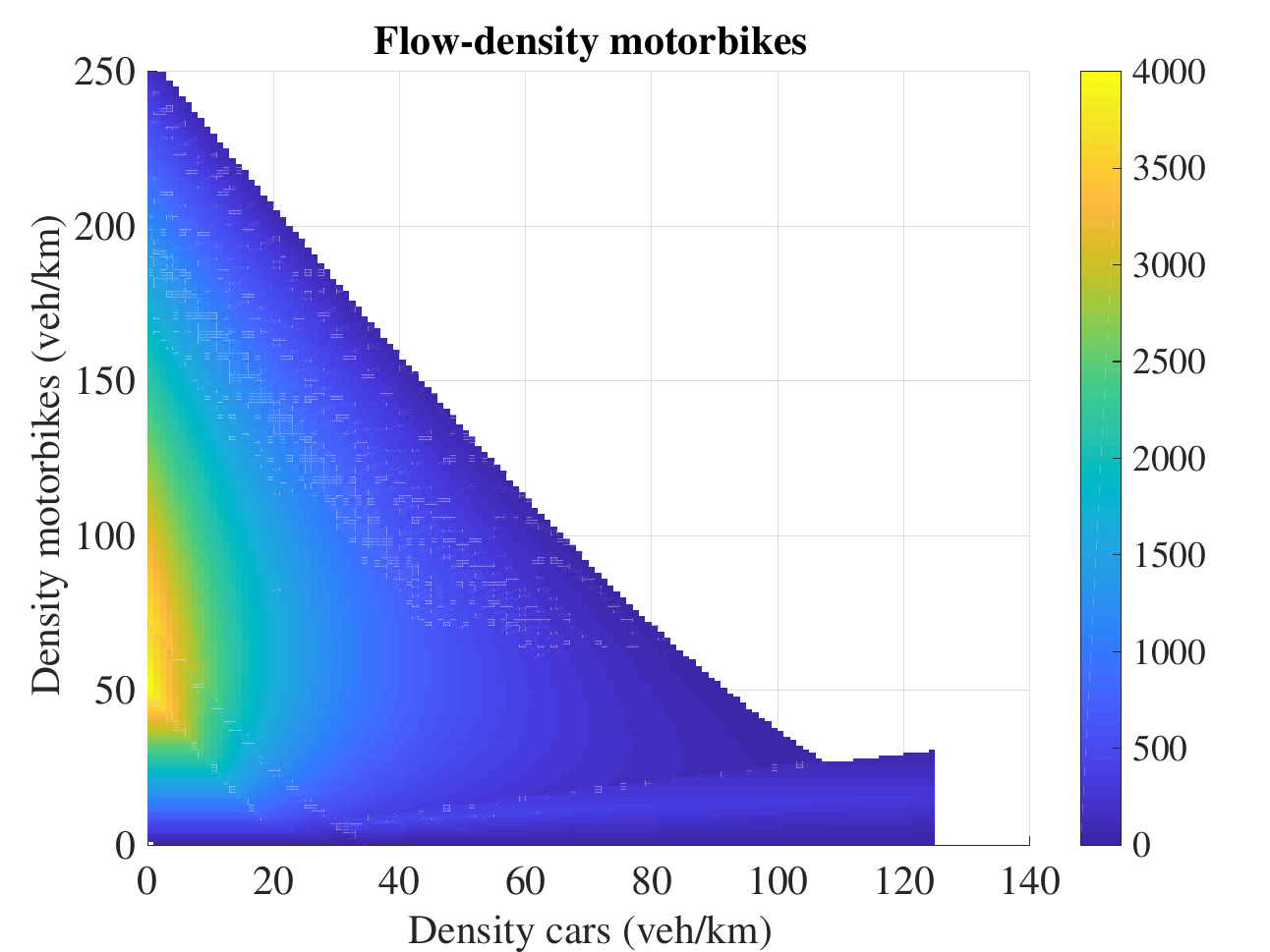}}
\caption{Fundamental diagrams}\label{fig_FDs}
\end{figure}

This way we find the final class specific fundamental diagrams. We numerically evaluate them and shwom them in \fig \ref{fig_FDs}.
Note how the lines of the phase transition between phase 2 and 3 is clearly visible. Note moreover how the speed of the motorbikes in phase 3 depends on the density of motorbikes (reduces with increase of motorbikes), but also with the density of the cars, since the limitation can be the speed difference with the cars. The total flow is shown as iso-flow lines in \fig \ref{fig_totalflow}.

\subsection{Capacity}
\begin{figure}
\subfigure[Iso flow lines]{\includegraphics[width=\halvepagina]{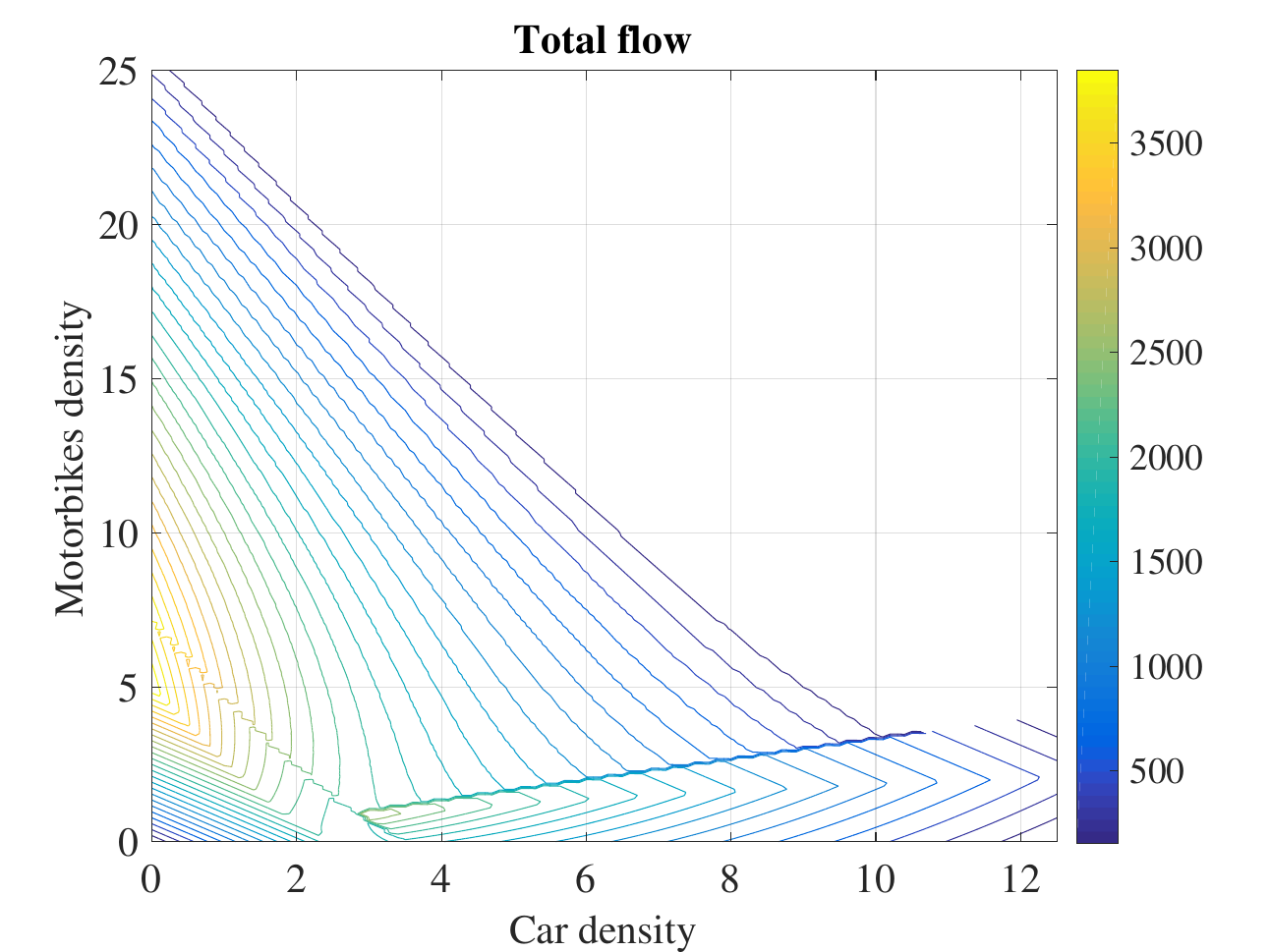}\label{fig_totalflow}}
\subfigure[Fraction of car traffic]{\includegraphics[width=\halvepagina]{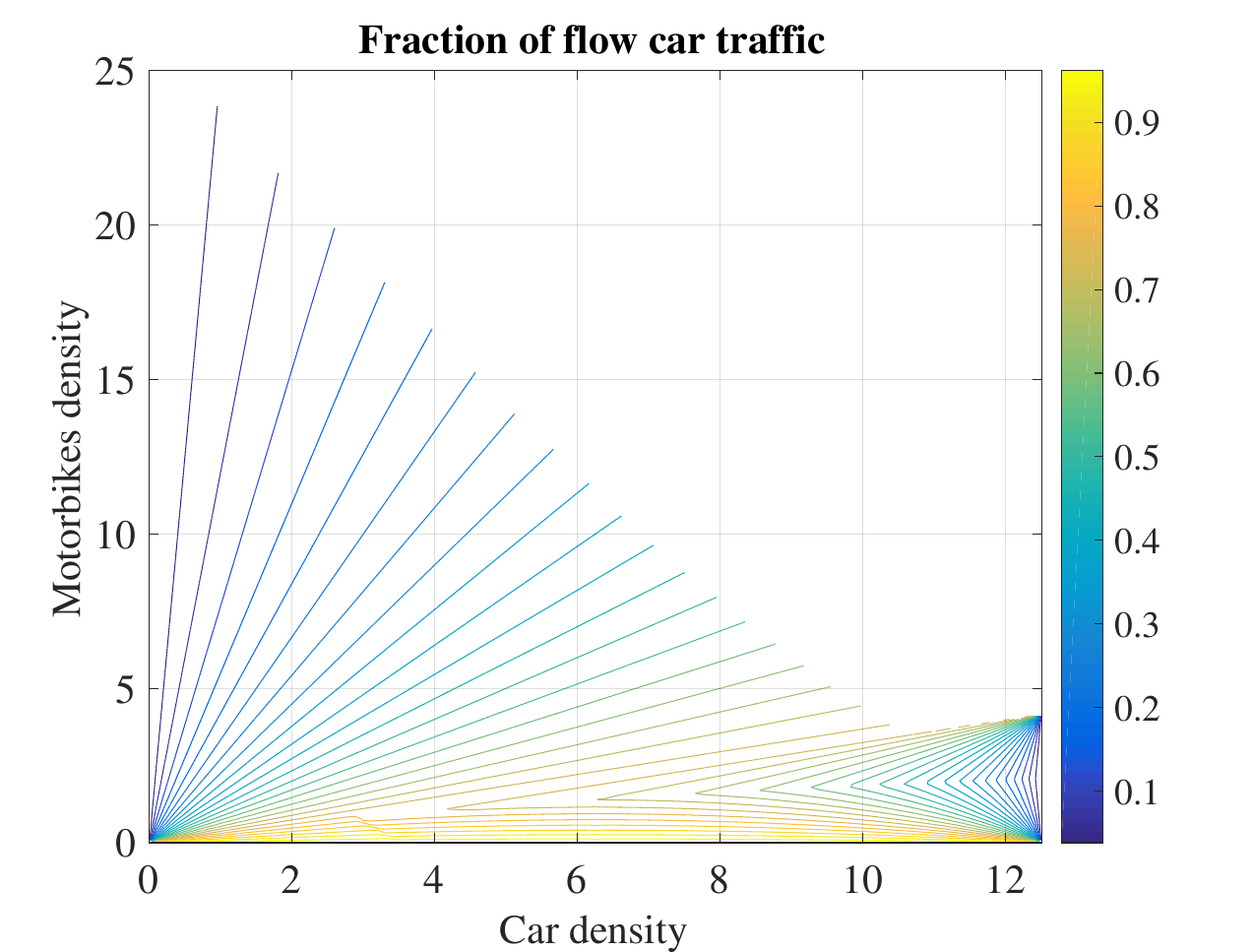}\label{fig_fraction_flow}}
\subfigure[Capactiy as function of car traffic]{\includegraphics[width=\halvepagina]{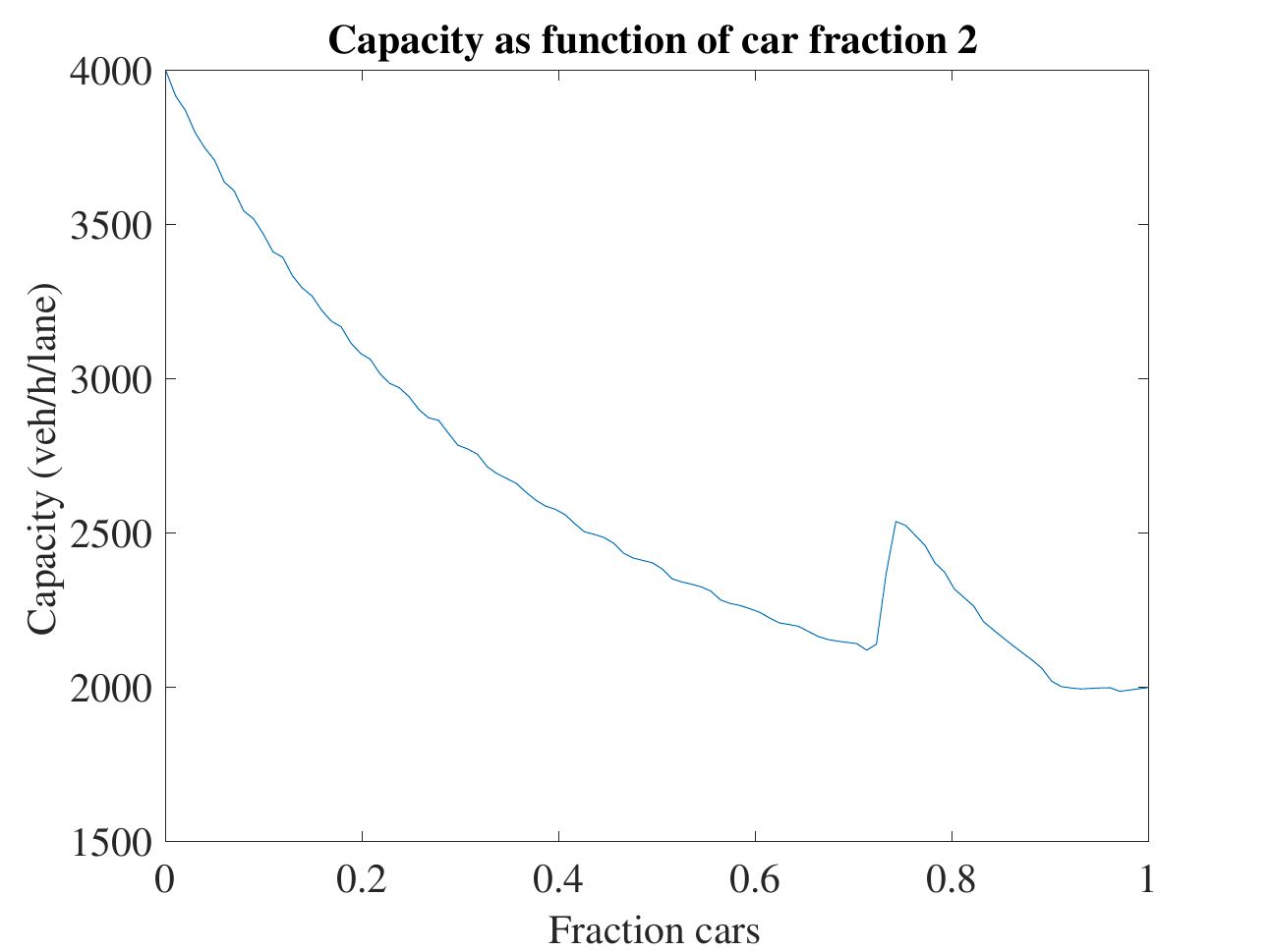}\label{fig_capacity_curve}}
\caption{Capacity}
\end{figure}

For traffic operations, capacity of the joint system is relevant. We hence add the flows of cars and motorbikes, and consider the total flow. Following the analyses of section \ref{sec_nexttoeachother}, this flow is a function of the density of both classes. This is shown in \fig \ref{fig_totalflow}. 

In this section we will determine the capacity as function of the fraction of motorbikes. Note that the fraction of motorbikes can be expressed as fraction of the flow and fraction of the density. From a practical point of view, one might consider the effects of certain fraction of drivers being stimulated to change to a motorbike. This means a certain fraction of the flow is changing modes, and hence we are considering a fraction of the flow. 

Figure \fig \ref{fig_totalflow} shows iso-flow lines if we add up flow of cars and motorbikes. As side note, straight lines from the origin represent the same ratio of density of motorbikes and density of cars. Since we are interested in the fraction of flows, the ``iso-fraction lines'' (i.e., lines over which the fraction of flows is equal) are different. They are shown in \fig \ref{fig_fraction_flow}. 

For each of this fractions, we find the point where the total flow is maximized. This is the capacity point for that fraction. The resulting capacity curve as function of fraction of cars is shown in \fig \ref{fig_capacity_curve}. (The curve is slightly fluctuating due to the numerical nature it has been created.) The extremes are not surprising: at 100\% car traffic, the capacity is 2000 veh/h as set in the fundamental diagram. At 0\% car traffic, and hence 100\% motorbike traffic, the capacity is 4000 motorbikes/h, which is caused by the fact that motorbikes can drive side-by-side, which hence doubles the capacity. It could be expected that the capacity gradually decreases with higher car fractions. However, at around 70\% car traffic, the graph shows another capacity than the other gradually decreasing trend. This is because the capacity state is in another phase. For the states with high fraction of cars, the capacity state is with all cars and motorbikes in capacity in the main lane. For low car fraction, the capacity state is with all main lanes at capacity. Recall that the filtering lane can only be used if the speed in the main lane is below 50 km/h. For these speed, the flow reduces and to compensate the flow in the filtering lane should be high. For very low fraction of motorbikes, there are simply not enough motorbikes to fill this lane. For higher fractions of motorbikes, the flow across all lanes reduces too much. However, at a moderately high motorbike fractions, the highest flows are found in a two-pipe regime, phase 3. In these conditions, the speed for motorbikes is higher than the speed for cars. 

\section{Dynamics}\label{sec_dynamics}
In this section we describe the dynamics of the model. We first indicate the way we solve the model, and then show the outcomes of the working in a case study. We also use the case study to extract some insights on the effect of the fraction of motorbikes. 

\subsection{Dynamic model}
As indicated in the introductions, most recurrent bottlenecks are at one location. From a modelling persepective, a dynamic model in Lagrangian coordinates (i.e., coordinates that move with traffic \citeh{Lec:2007-1, Lec:2007-2}) might be preferred. This has already been developed \citeh{Wie:2019}. However, from a practical perspective on applying this to a bottleneck (fixed in space), this is not convenient to work with. Vertical queuing approaches are not useful either. If chosen, this would (1) not provide information on the location of the queue, and (2) not do justice to the delicate balance in speeds in the queue between the classes. Therefore, we rather prefer a macroscopic model in which cells which remain fixed in space. 

As input to this model we use the traffic demand at the beginning of boundary 1, as well as capacity constraints (lane restrictions) inside the model. The model should then predict the future traffic conditions. The hard part is that there are two pairs of fundamental diagrams -- i.e., both classes compete for the same scarce space. For this, a generic solution is available.

Let's consider system from a more formal mathematical view. We have a conservation of vehicles for each class, so in equations:
\begin{equation}
\begin{cases}  \dfrac{\partial \dens_c}{\partial t} +
 \dfrac{\partial \dens_c v_c(\dens_c,\dens_m)}{\partial x} &=0\\
\dfrac{\partial \dens_m}{\partial t} + \dfrac{\partial \dens_m v_m(\dens_c,\dens_m)}{\partial x} &=0\end{cases}
\end{equation}
This can also be expressed in a vectorial form:
\begin{equation}
\dfrac{\partial V}{\partial t} + A(V)\dfrac{\partial V}{\partial x} = 0,
\end{equation}
with $V=\begin{pmatrix}\dens_c, \dens_m \end{pmatrix}$.  Then, $A(V)$ is
\begin{equation}
A(V)=
\begin{pmatrix}
v_c+\dens_c\dfrac{\partial v_c}{\partial \dens_c} & \dens_c \dfrac{\partial v_c}{\partial \dens_m}\\
\dens_m \dfrac{\partial v_m}{\partial \dens_c} &v_m+\dens_m\dfrac{\partial v_m}{\partial dens_m}
\end{pmatrix}
\end{equation}
For reasons of convenience, and reference to hyperbolic papers, we simplify the notation to:
\begin{equation}
A(V)=
\begin{pmatrix}
\beta_1 & \alpha_1 \\
-\alpha_2 & \beta_2\end{pmatrix}
\end{equation}

The eigenvalues of A(V) give the (characteristic) wave speeds. For the systems as presented, we have the following two eigenvalues
\begin{equation}
\begin{cases}\lambda_1&=\dfrac{1}{2}\left(\beta_1+\beta_2-\sqrt{D}\right)\\
\lambda_2&=\dfrac{1}{2}\left(\beta_1+\beta_2+\sqrt{D}\right)\end{cases}
\end{equation}
with $D=(\beta_1-\beta_2)^2+4 \alpha_1 \alpha_2$

We aim to get the flow at a boundary between two cells. At this location, there will be 4 speeds which might be relevant, being $\lambda_1$ and $\lambda_2$ from the upstream cell (denoted $\lambda_1^g$ and $\lambda_2^g$) and $\lambda_1$ and $\lambda_2$ from the downstream cell, denoted $\lambda_1^y$ and $\lambda_2^u$. We now take the minimum and maximum of these 4 speeds, and call them $Su$ and $Sd$. Note this is a default name giving within a HLL scheme, and the $u$ and $d$ do not refer to the downstream and upstream cell.
\begin{eqnarray}
Su&=\min\left(\lambda_1^g;\lambda_2^g;\lambda_1^u;\lambda_2^u\right)\\
Sd&=\max{\lambda_1^g;\lambda_2^g;\lambda_1^u;\lambda_2^u}
\end{eqnarray}
A physical interpretation is that $Sd$ is the speed of the fastest traffic stream and $Su$ is the wave speed with the lowest speed. Hence, $Sd>0$, and the sign of $Su$ depends on the traffic conditions. 

We now aim to determine the flow over the boundaries, $Q$ (note we look for 2 flow values, one per class). The flows inside the cells are denoted $Qu$ for the upstream cell and $Qd$ for the downstream cell. We determine the two factors, $Su$ and $Sd$. $Su$ is the derivative of the flow to (total) density at the densities in the upstream and downstream cell. $Sd$ is the maximum of the speed of both classes in both the upstream and downstream cell (i.e., maximum of 4 numbers). 

A key aspect in our computation scheme, and important for the accuracy is that we determine $Su$ and $Sd$ \emph{for each time step and for each boundary}. Hence we do not globally see the ranges possible for $Su$ and $Sd$, but we stay as close as possible to the values at that particular boundary. Numerically, this is a bit more time consuming, but with derivatives for the speeds determined on beforehand, this is a look-up operation which can be easily performed.

We now use the  Harten-Lax-van Leer (HLL) Riemann Solver \citeh{Har:1983}. The principle is that only the fasest and slowest wave are being considered  \citeh{Tor:1994,Tor:2013}, which we find from $Su$ and $Sd$. Using these, we can define a $Q_\textrm{HLL}$:
\begin{equation}
Q_\textrm{HLL}=\frac{
Qu*Sd-Qd*Su+
Sd*Su*\left(Kagg_\textrm{downstream}-Kagg_\textrm{upstream}\right)}{Sd-Su}
\end{equation}
Note that this equation weights the flows upstream and downstream with $Su$ and $Sd$.

Then, finally, we find the flow over the boundary b, depending on the values of $Su$ and $Sd$:
\begin{equation}
Q=\begin{cases} 
Qu& \textrm{if} Su>=0 \\
Qd& \textrm{if} Sd<=0\\
Q_{HLL} &\textrm{otherwise}
\end{cases}
\end{equation}

For the flow at the boundaries, we use the demand as inflow (externally ensuring there is no congestion at the first cell), and the flow in the last cell as outflow (ensuring there is no downstream congestion by choosing the bottleneck within the simulated section)

\subsection{Illustration of the working}
This section shows how the model works. We first present the case we present, followed by the resulting traffic operations for the case. We will then bring the results to a more generic level and discusses the generic patterns. 
\subsubsection{Case set up}
\begin{figure}
\includegraphics[width=\halvepagina]{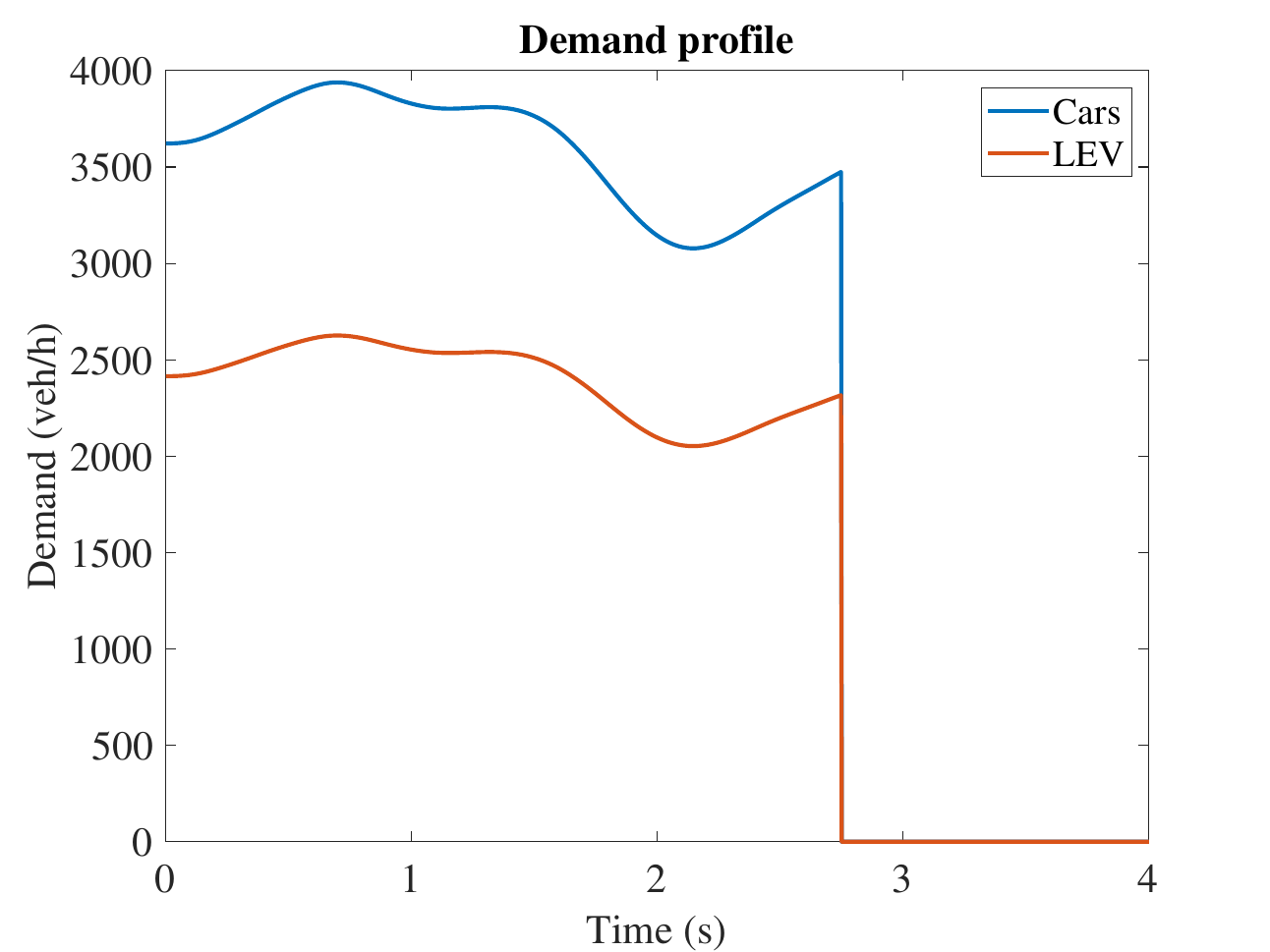}
\caption{Demand profile: for 60\% car traffic and 40\% motor motorbike traffic}\label{fig_demand}
\end{figure}

For the case study, we consider a road with a lane drop bottleneck. The 4 lane section becomes a 3 lane section at which goes from 4 lanes. The demand is modelled on the morning peak demand of the A4 northbound motorway near Delft. After 2.5 h, the demand is reduced to 0 and the network can empty. We take the total demand and consider a fraction of that demand is transferred to motorbikes. The fractions considered are: [1 0.99 0.98 0.97 0.96 0.95 0.9 0.8 0.7 0.6 0.5 0.4 0.3 0.2 0.1 0]. An example of the demand profile is shown in \fig \ref{fig_demand}.

We choose a time step of 10 seconds, and a (free-flow speed matching) cell length of 222 meters. The total network is 8,000 m and the lane drop is at 6,400 m.

\subsubsection{Model results}
\begin{figure}
\includegraphics[width=\derdepagina]{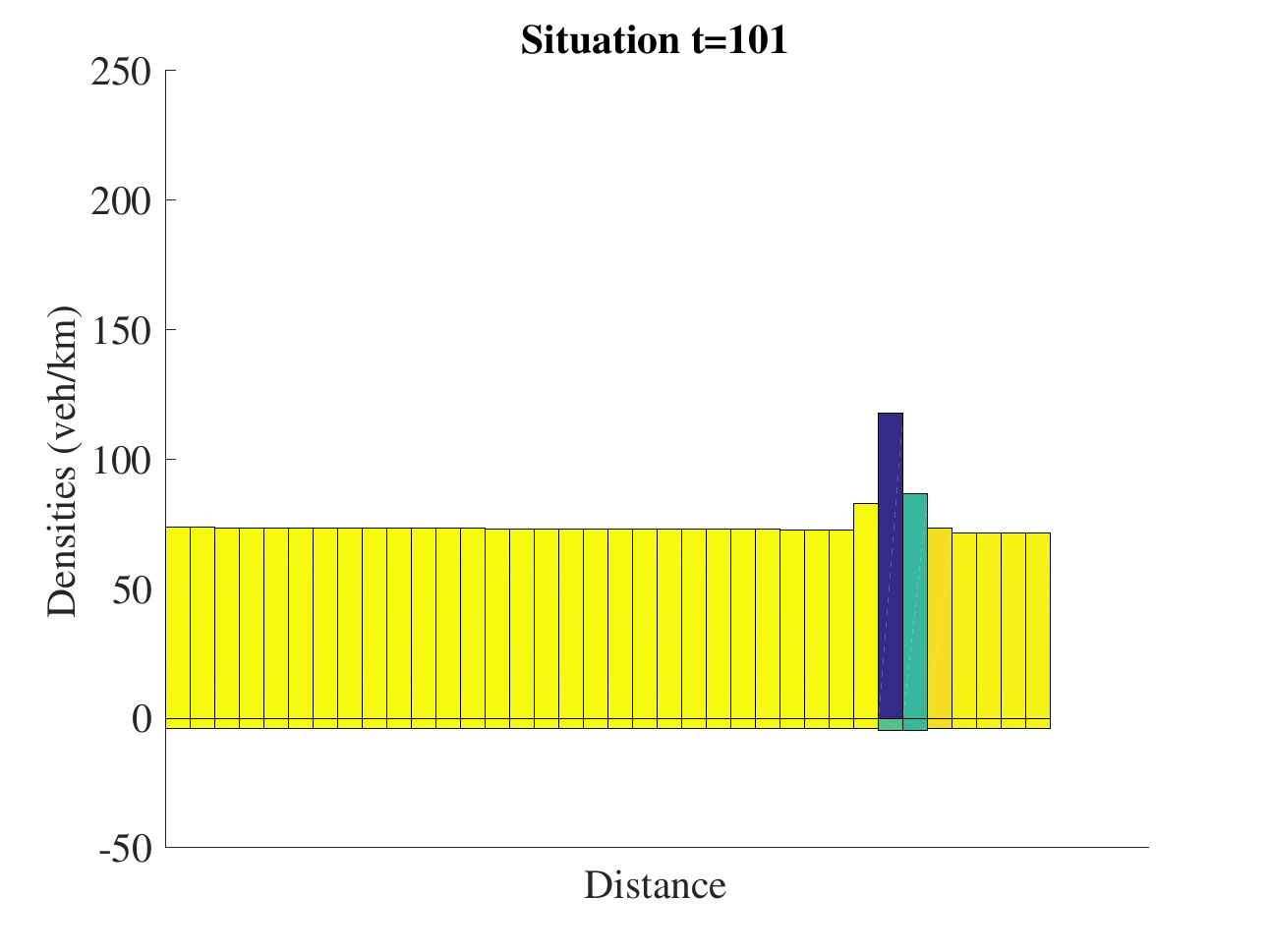}
\includegraphics[width=\derdepagina]{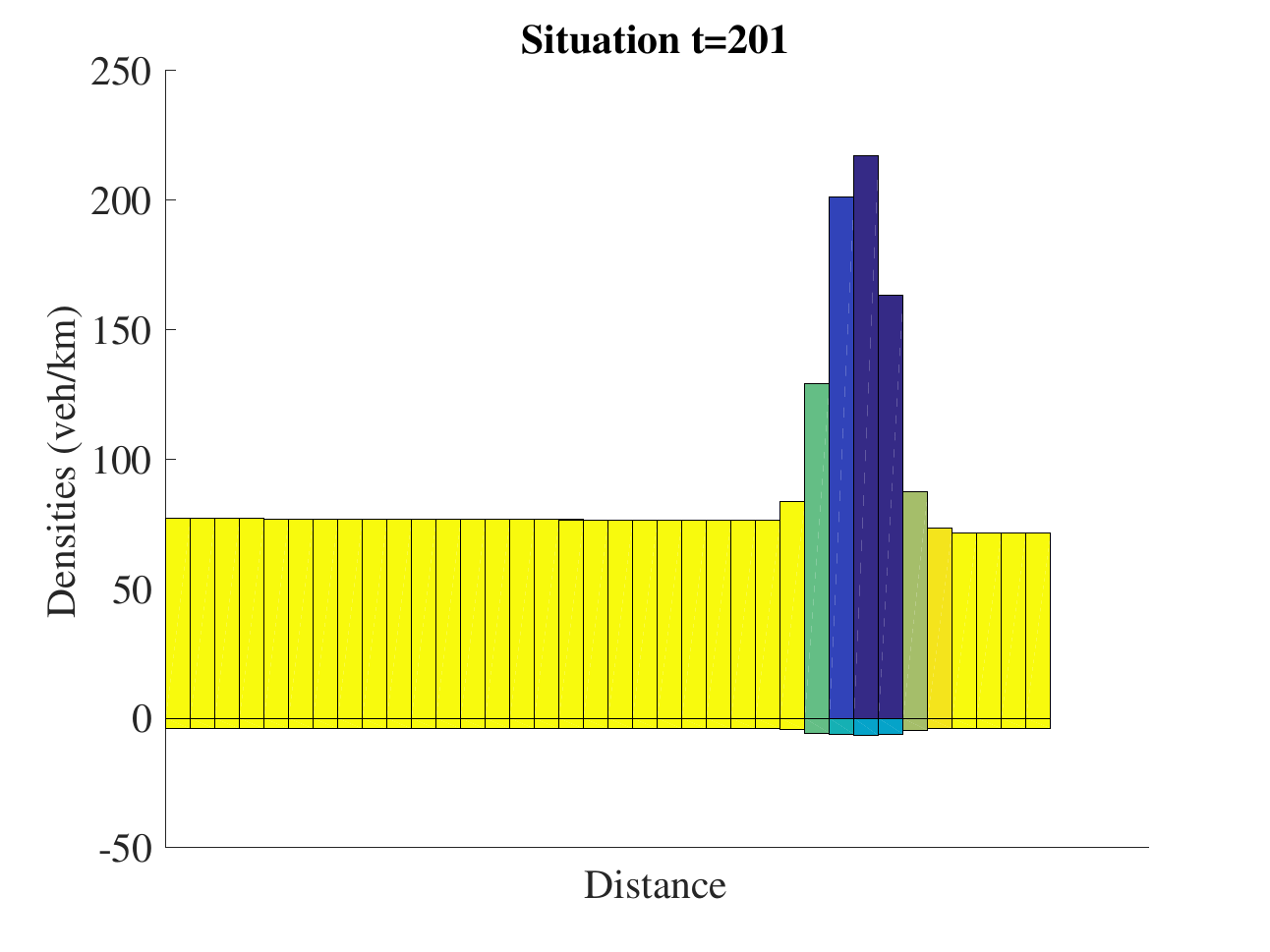}
\includegraphics[width=\derdepagina]{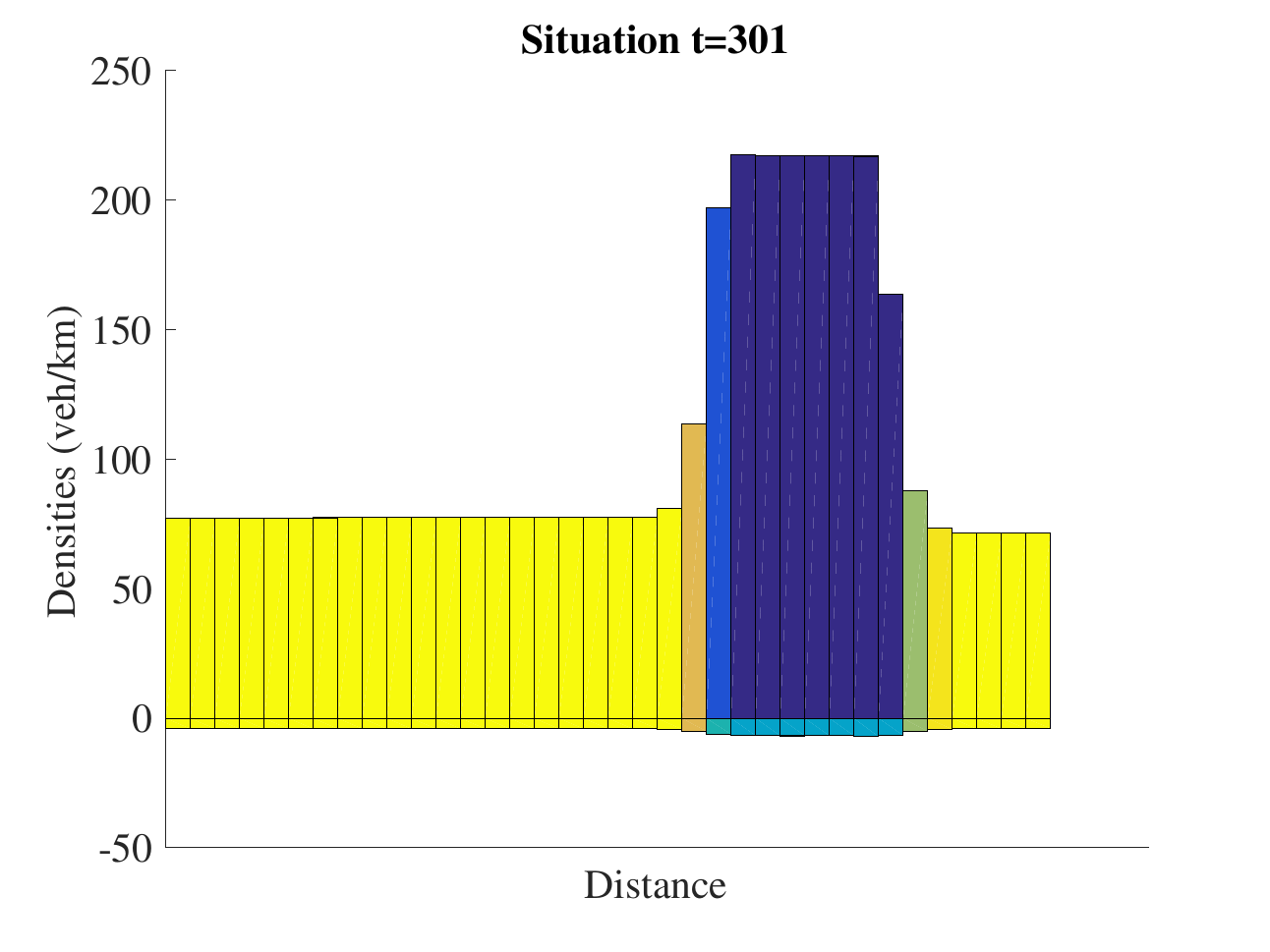}\\
\includegraphics[width=\derdepagina]{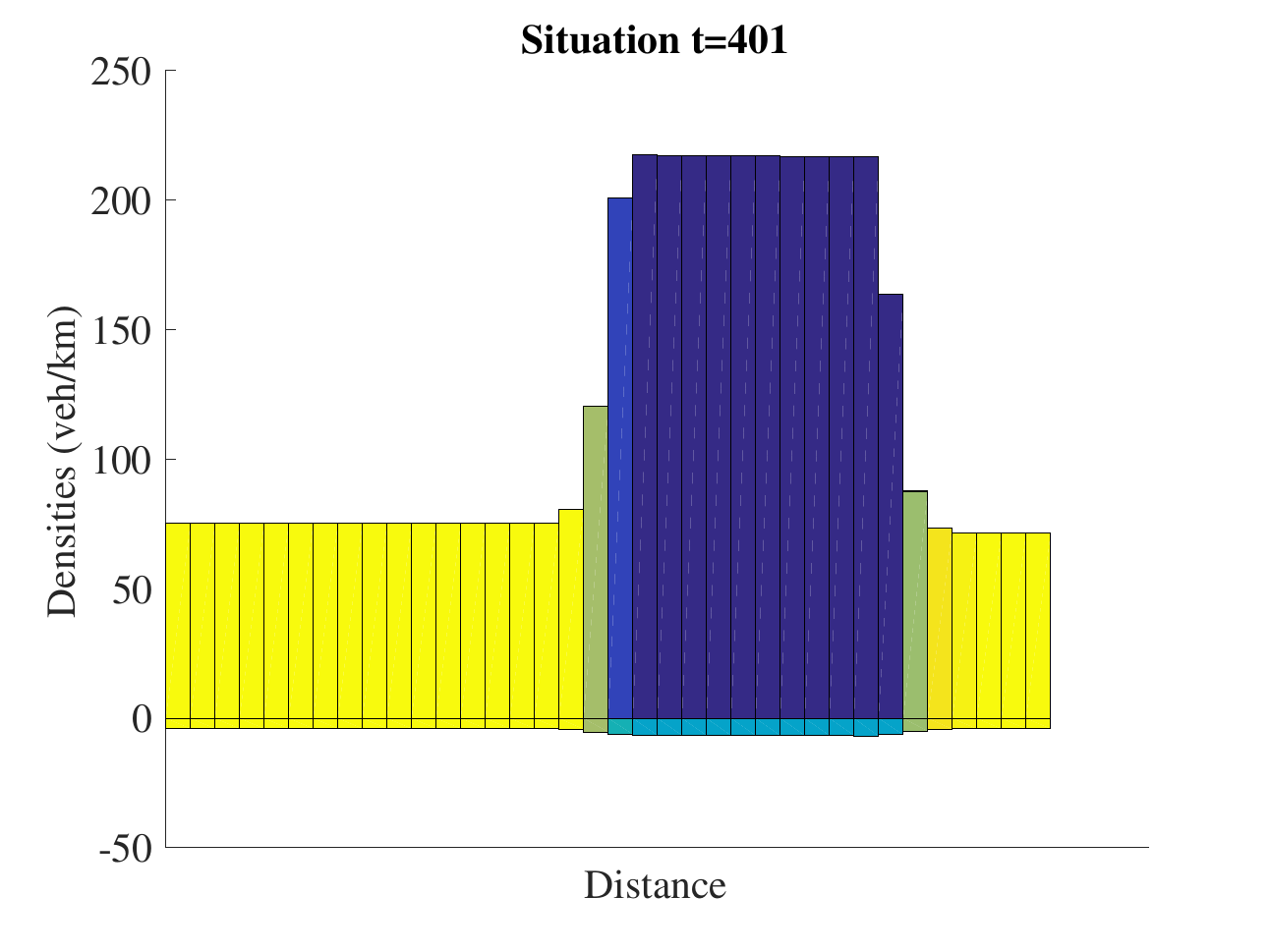}
\includegraphics[width=\derdepagina]{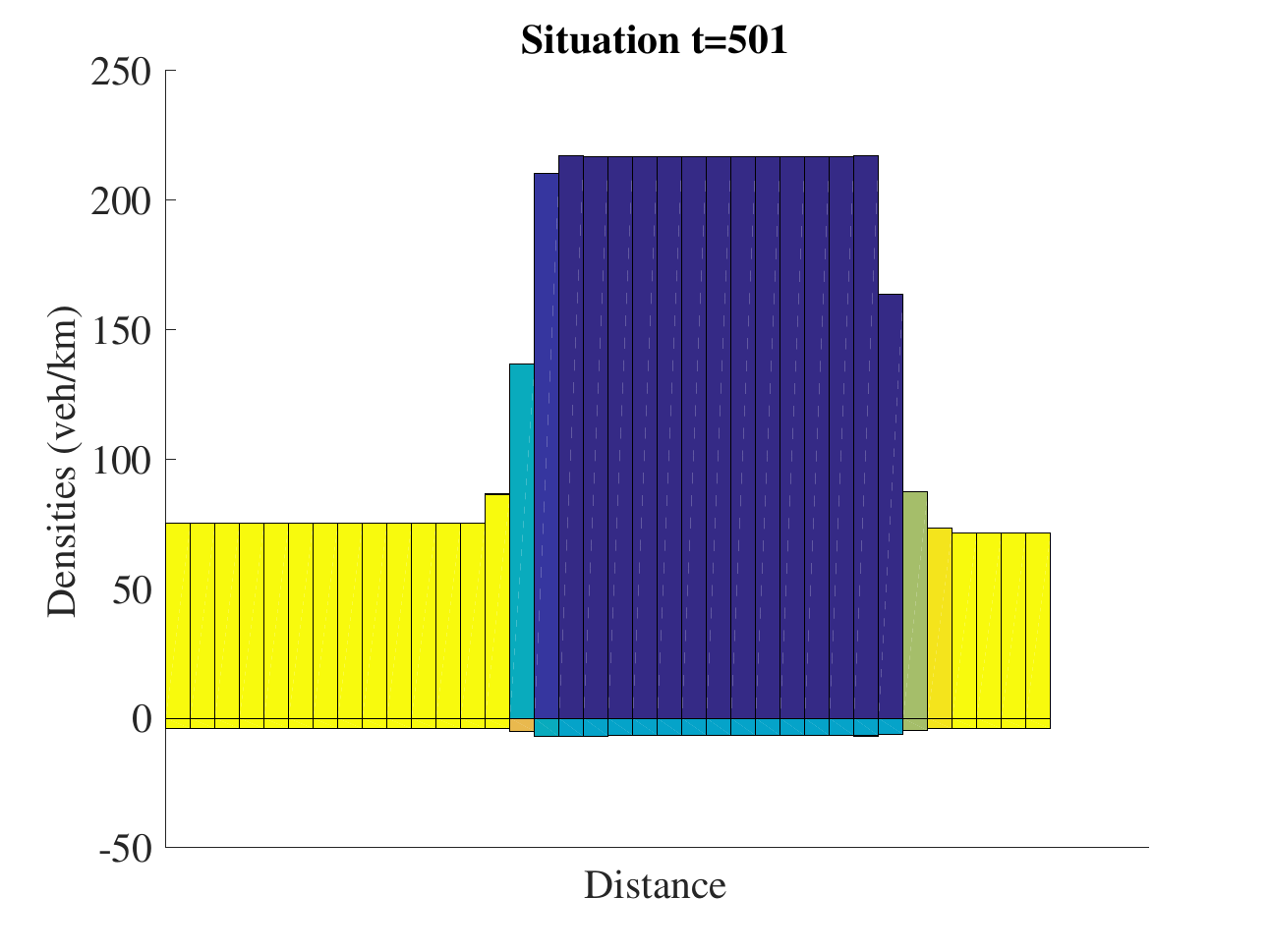}
\includegraphics[width=\derdepagina]{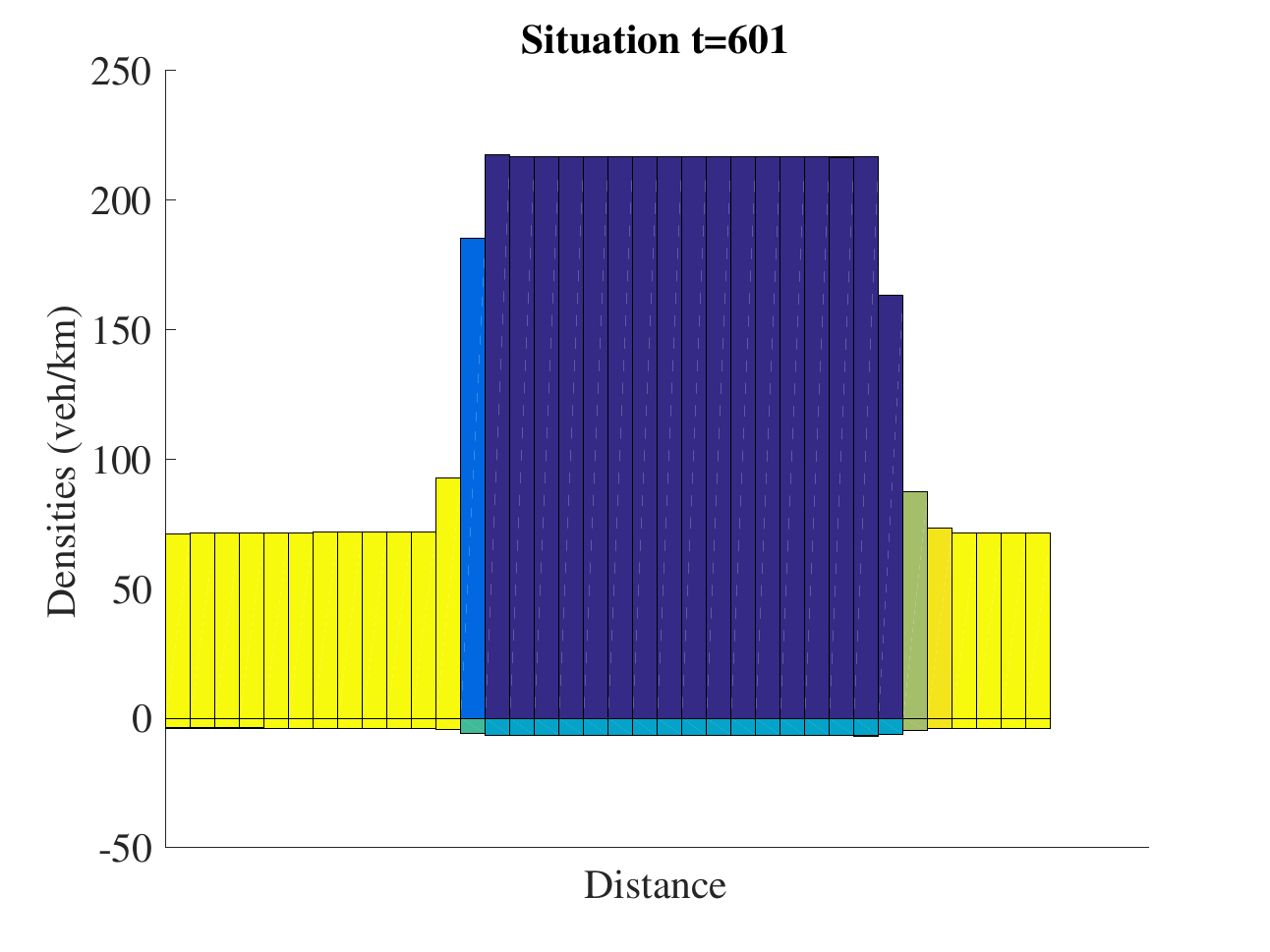}
\caption{Traffic states at various time instances. Hight indicates density (cars to positive y-direction, motorbikes to negative y-direction). The color of the (class-specific) bar indicates (class-specific) speed}\label{fig_trafficstates}
\end{figure}

\Fig \ref{fig_trafficstates} shows the densities and speeds of the two classes for various time instances. For reasons of convenience, in our simulation we used not the tightest bounds for the $Sd$ and $Su$; this induces some numerical errors, but overall it does suffice to show the impact of the influence of motorbikes and the resulting traffic situation. For both classes, the head of the jam is at the same location (namely at the bottleneck location), and it starts at the same time; also the tail is at the same location. This can be understood from a physical point of view since regardless of the fraction of motorbikes, they need to slow down whenever the speed of cars is low. The speed of the motorbikes in the queue is different from the speed of the cars. This is also understandable from the point of view of the fundamental diagrams, and from physical interpretation. Motorbikes can move in between the queued cars. 

\begin{figure}[h]
\subfigure[Travel times as function of departure time at a 5\% motorbike fraction]{\includegraphics[width=\halvepagina]{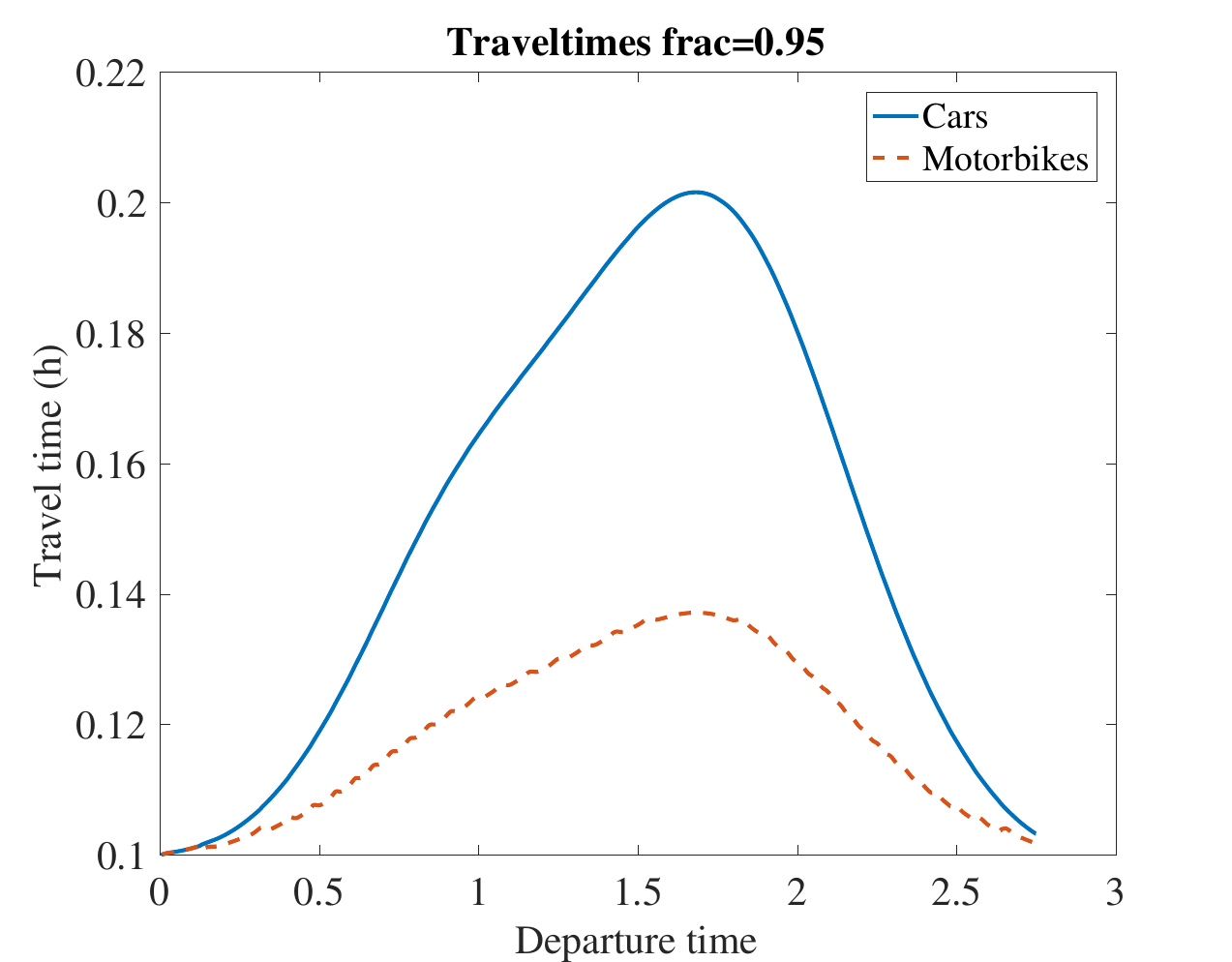}\label{fig_traveltimes}}
\subfigure[Overall travel times]{\includegraphics[width=\halvepagina]{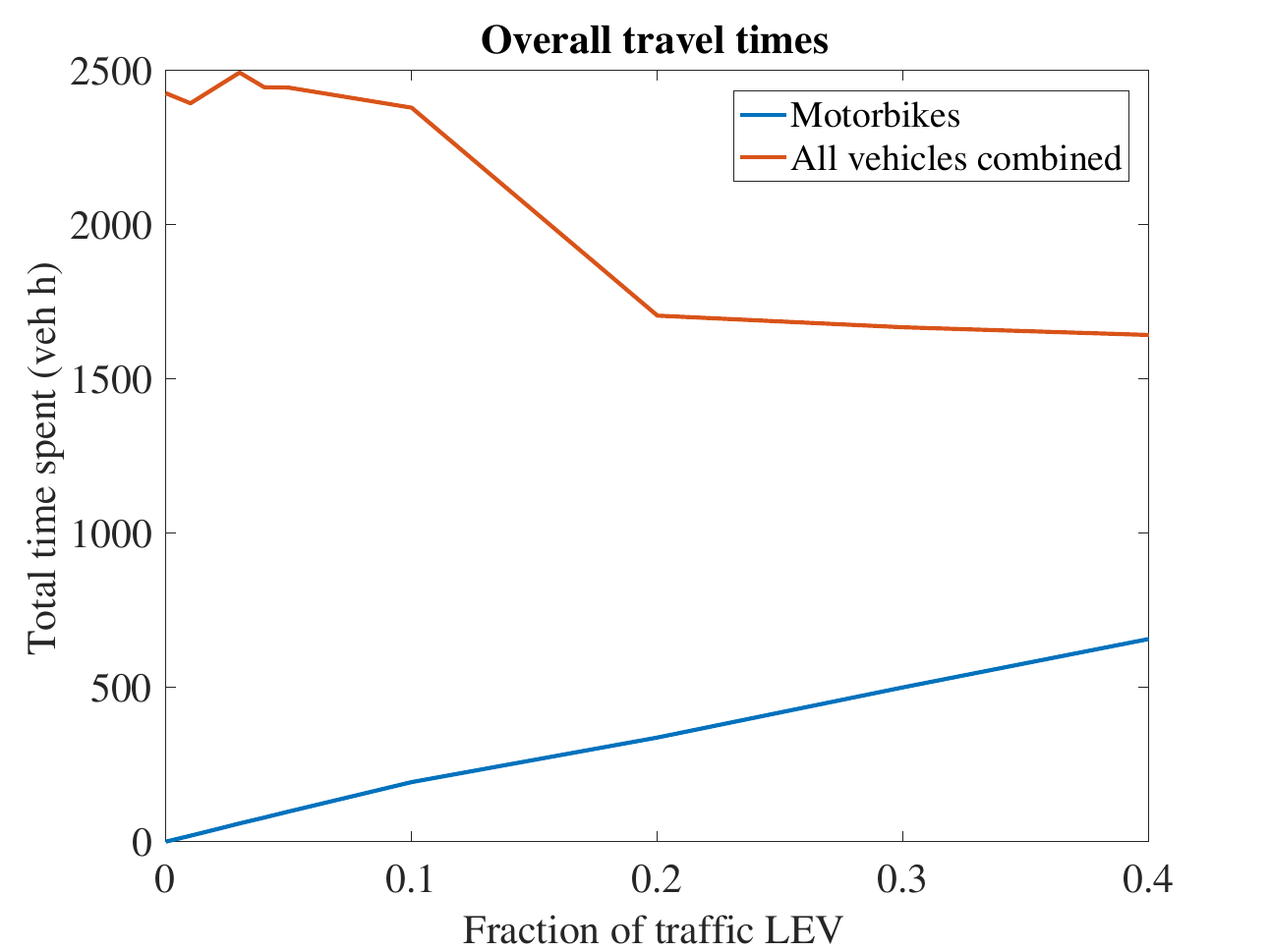}\label{fig_reistijd_tot}}
\subfigure[Mean travel times as function of motorbike fraction]{\includegraphics[width=\halvepagina]{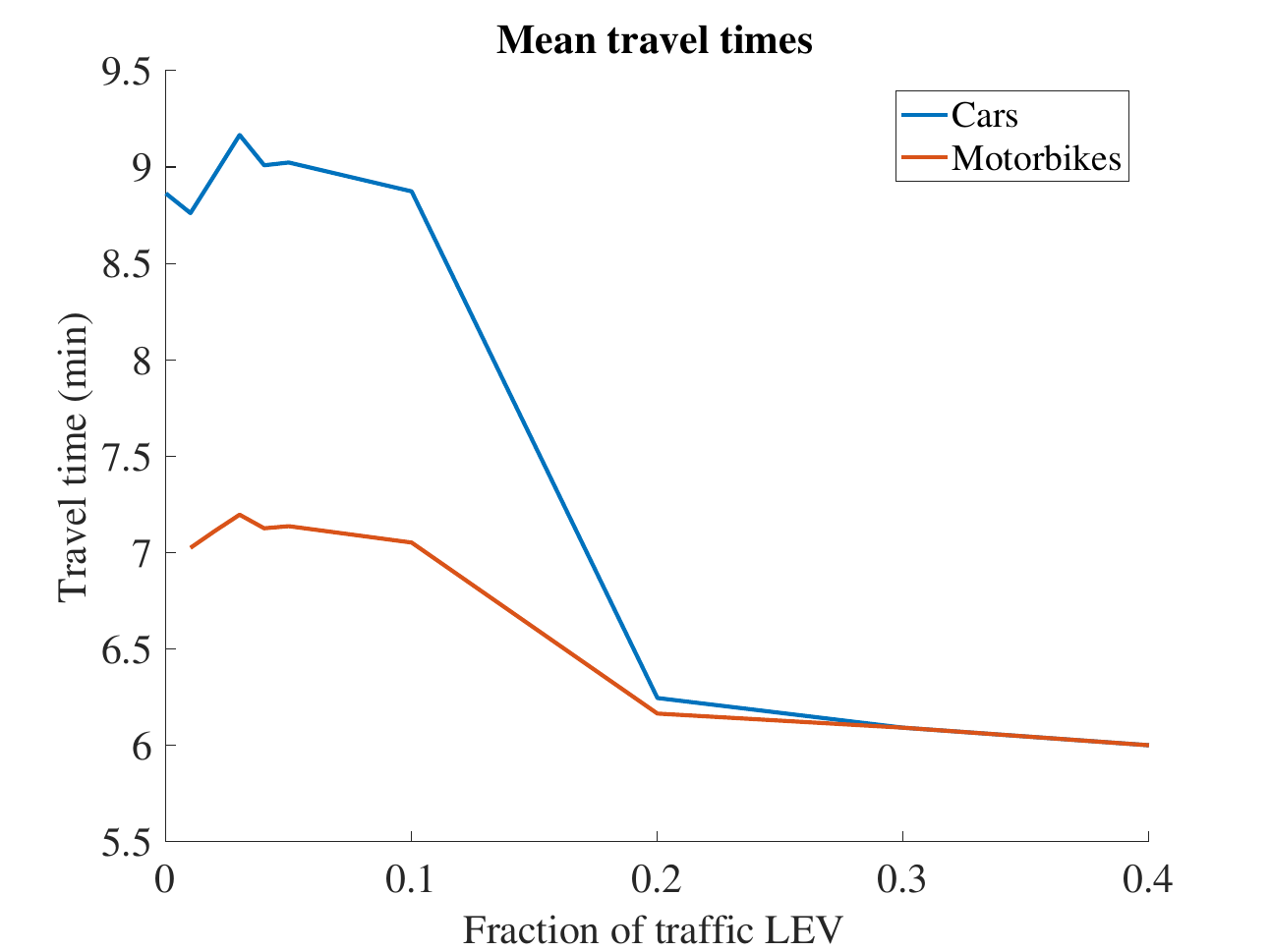}\label{fig_ttfractionmb}}
\caption{Travel times}
\end{figure}

The dynamics lead to class-specific travel times for particular departure times. Note we present here the full travel times for one vehicle, and not the instanteneous travel times (i.e., the travel time if a vehicle would experience all the class-specific speeds as measured at one moment in time). These are shown in \fig \ref{fig_traveltimes}, for a 5\% fraction of motorbikes. They increase and decrease both in a similar way, but at different speeds. Note that a base line in the figure, a 0.1h travel time, is the free flow travel time for the segment. 

\subsection{Insights on effects of fraction of motorbikes}\label{sec_insights}
The total time spent can be found by taking the weigted sum of the travel times per class. This is shown in \fig \ref{fig_reistijd_tot}

Weighting this figure with the amount of vehicles experiencing this travel time, we can assess the mean travel times per class as function of the fraction of motorbikes in the traffic stream. This is depicted in \fig \ref{fig_ttfractionmb}. There we see that higher motorbike percentages generally lead to lower travel times, as expected. Surprisingly, the travel times for motorbikes and cars are the same from around 25\% fraction of flow as motorbike traffic. In the considered case, the travel times have almost reduced to the free flow travel time. Interestingly, this also holds in case of even higher demands. 

For instance, let's take the thought experiment of a case with much higher flows and higher motorbike fractions ($> 25\%$). There will be congestion (since there are higher flows), but since the motorbike fraction is high, we will see a congested state in phase 1 (see \fig \ref{fig_phases}). In fact, traffic slows down and there will be motorbikes in the filter lane. However, not all motorbikes will fit in the filter lane, hence there are some motorbikes in the main lines. Therefore, they will choose the fastest lane of the filter lane and the main line, which therefore will operate at the same speed. Since both types of lane operate at the same speed, the type of vehicles will also move at the same speed. 

\section{Discussion and conclusions}
This paper has introduced a model for multiple vehicle classes. In addition to earlier works, it has explicitly considered traffic operations that is partially bound to the main lines, and partially not when motorbikes can filter through the lanes. Based on these assumptions, several phases of traffic were derived. Also fundamental diagrams were developed, one for both classes with density of both classes as independent variables. The capacity was increasing with the fraction of motorbikes. Surprisingly, the phase of capacity point depends on the  fraction of motorbikes: this can be a 1 pipe or 2 pipe regime. 

A numerical scheme to solve this system has been presented, as well as a case study in which we show the traffic dynamics. This approach has shown that the model can be solved with a cell based flow scheme, using dedicated techniques adapted from numerical mathematics. The numerical scheme itself is useful and can be adapted to any combinations of classes in a macroscopic traffic model.

Application to a case study showed how to model can illustrate the speeds and travel times for two classes. Increasing the motorbike fraction decreases travel times for both fractions, since the capacity increases. For this reason, a policy could be to increase the motorbike fraction. At lower motorbike fractions, there is an individual incentive to change to a motorbike too: the travel time is lower because they can filter through congestion. For higher fractions of motorbikes, the collective gain of a change from car to motorbike still remains present (i.e., collectively, all travellers are better of if someone changes from car to motorbike), but the comparative advantage disappears (i.e., the travel time gain from a mode shift is the same for the person changing from car to motorbike as for someone else not changing).  In this case, additional measures to stimulate shifting are more likely needed.

\bibliographystyle{plainnat}
\bibliography{20200730}
\end{document}